\documentclass[12pt,twoside]{article}
\usepackage{fleqn,espcrc1}

\usepackage{graphicx}
\usepackage[figuresright]{rotating}

\newcommand{\AmS}{{\protect\the\textfont2
  A\kern-.1667em\lower.5ex\hbox{M}\kern-.125emS}}

\title{Effect of nuclear structure on Type Ia supernova nucleosynthesis}

\author{D.J. Dean\address{Physics Division, Oak Ridge National Laboratory, \\
        P.O. Box 2008, Oak Ridge, Tennessee 37831-6373 USA}
        \thanks{This research was sponsored by the Division of
         Nuclear Physics of the U.S. Department of Energy under Contract
         No. DE-AC05-00OR22725 managed by UT-Battelle, LLC.}}
\begin{document}

% typeset front matter
\maketitle

\begin{abstract}
The relationship among nuclear structure, the 
weak processes in nuclei, and astrophysics becomes quite
apparent in supernova explosion and 
nucleosynthesis studies. 
In this brief article, I report on progress made in 
the last few years on calculating electron
capture and beta-decay rates in iron-group nuclei. 
I also report on applications
of these rates to Type-Ia nucleosynthesis studies. 
\end{abstract}

\section{Introduction}

Nuclear physics plays an important role
in stellar evolution. 
Fusion, proton capture, and neutron capture
are all examples of the importance of the electromagnetic interaction
in creating energy that powers stars. 
Weak interactions in nuclei, including electron capture and
$\beta$-decay, play an important
role in the evolution of both Type Ia and II supernovae and
their nucleosynthesis. 
In this brief article, I will 
discuss recent progress in understanding the nuclear 
physics involved in Type Ia explosion mechanisms.
I focus on progress made in accurately calculating 
electron-capture and $\beta$-decay rates in iron group nuclei.  

In order to understand weak processes in nuclei, it becomes
necessary to properly describe the nuclear structure of
the relevant systems.  Short of a complete solution to 
the many-body problem, the shell model is widely
acknowledged to be the appropriate theoretical tool to 
describe both ground- and excited-state properties of nuclei. 
The shell-model requires as input a reasonable valence model space
and a reliable effective two-body interaction that reproduces
known properties of nuclei within the given model space. Such
interactions exist for $p$-, $sd$-, and $pf$-shell 
nuclei, and are under development for heavier or more neutron-rich 
systems. In this paper, I discuss
calculations made in the $pf$-shell using the two-body effective
interaction KB3 \cite{poves}, or slight modification thereof.   

The very nature of the quantum many-body problem for fermions -- its 
inherent computational difficulty due to the necessary antisymmetrization
of the many-body wave-function -- requires significant computational 
capability and expertise. This is particularly true of approaches that
are trying to treat the many-body problem exactly, 
or in extremely large shell-model spaces.
Standard shell-model diagonalization techniques
have recently progressed into the $pf$-shell \cite{gabriel2}
while other techniques based on Monte Carlo algorithms
have also been quite successful in recent years \cite{koonin}.

\section{Electron capture in Type Ia supernovae}

Electron capture on nuclei 
takes place in high density matter where the Fermi energy
of a degenerate electron gas is sufficiently large
to overcome the energy thresholds given by the negative Q-values of 
such reactions.  
Type Ia supernovae appear to be thermonuclear explosions of 
white dwarfs in binary systems. The high accretion rates of the 
white dwarf permits relatively
stable H- and He-shell burning and leads to a growing C/O white dwarf. 
When the white dwarf mass is sufficiently close to the Chandrasekhar mass, 
gravitational contraction sets in, and the
central density becomes high enough to ignite
carbon fusion. The environment 
of a degenerate electron gas provides a pressure that depends only
on the density; therefore, the initial heat generation does not lead to
the pressure increase and expansion that would lead to a controlled and
stable burning. Instead, a thermonuclear 
runaway occurs. The burning front propagates through the whole star, causing
complete disruption without a remnant. 

The high Fermi energy of the degenerate electron gas in the white
dwarf leads to efficient electron capture on nuclei in 
the high density burning
regions and reduces $Y_e=\langle Z/A\rangle$, the
electron fraction, or equivalently, the average proton-to-nucleon ratio,
during explosive burning in the center. This important factor
controls the isotopic composition ejected
from such explosions. Thus one test of theoretical models 
is whether they reproduce the observed isotopic compositions. 
If the central density
exceeds a critical value, electron capture can cause a dramatic reduction
in the pressure of degenerate electrons and can therefore induce collapse 
of the white dwarf. Thus, electron capture on intermediate-mass and
Fe-group nuclei plays a crucial role for the burning front propagation in 
Type Ia supernovae. Beta-decay is also relevant when the $Y_e$ values
correspond to $Z/A$ ratios of nuclei that are more neutron-rich than
stability.

Electron capture on nuclei in the energy regime relevant 
in Type Ia environments occurs primarily through
Gamow-Teller transitions. These transitions connect
an initial nuclear state to final states through 
a spin-isospin operator $\vec{\sigma}\tau_+$. 
The Gamow-Teller strength distributions enter directly
into the rate calculations for electron-capture by folding 
with the electron energy distribution in the stellar environment
of interest. 

The centroid of the Gamow-Teller strength distribution in
nuclei resides at several MeV of excitation energy in the daughter
systems. From the mid-1980s, this distribution was modeled
as a single matrix element at a given excitation energy,
carrying the total strength calculated from the non-interacting
shell model \cite{FFN}. These electron capture and $\beta$-decay
rates, known as Fuller, Fowler, and Newman (FFN) rates, 
are commonly used in astrophysical calculations. 
About six years ago, research using the 
shell-model Monte Carlo (SMMC) technique indicated that 
two ingredients were necessary to obtain Gamow-Teller results
that were compatible with experimental data: the complete 
0$\hbar\omega$ model space needed to be used, and 
calculations had to be performed using a 
reliable interaction for the model space of interest. 
If these two conditions are met, then one can reliably
reproduce experimental total Gamow-Teller strengths 
to within a constant factor \cite{kl95} and the 
strength distributions \cite{radha97}. Systematic deviations
from the FFN parameterization of the Gamow-Teller strength
were observed in the SMMC calculations.
In particular, for even-even nuclei the strength was found
to be lower than estimated by FFN, and in odd-A and odd-odd
nuclei the centroid of the strength was significantly higher 
than estimated by FFN \cite{dean98}. 

Recently, stellar electron capture and $\beta$-decay rates were
calculated by shell-model diagonalization using a slightly
improved version of the $pf$-shell two-body interaction 
for all $pf$-shell nuclei \cite{gabriel2} and confirm the 
trend already observed in SMMC studies. Systematic deviations from
the Gamow-Teller parameterization assumed by the FFN compilation lead to 
significantly smaller electron capture rates on odd-A nuclei and odd-odd
nuclei, and slightly smaller capture rates on even-even nuclei than
FFN. 

Using Gamow-Teller strengths generated from either SMMC calculations
or from the shell-model diagonalization, we implemented
new electron capture and beta-decay rates into 
Type Ia evolution codes \cite{brach00}. 
We compared the FFN and shell-model rates for several typical nuclei
and, based on this comparison, multiplied the FFN electron capture
rates by 0.2 (for even-even nuclei), 0.1 (odd-A), and 0.04 (odd-odd),
while the FFN $\beta$-decay rates were scaled by 0.05 (even-even),
0.025 (odd-A), and 1.7 (odd-odd). 

\begin{figure}
\begin{center}
\hbox{\includegraphics[scale=0.325,angle=90]{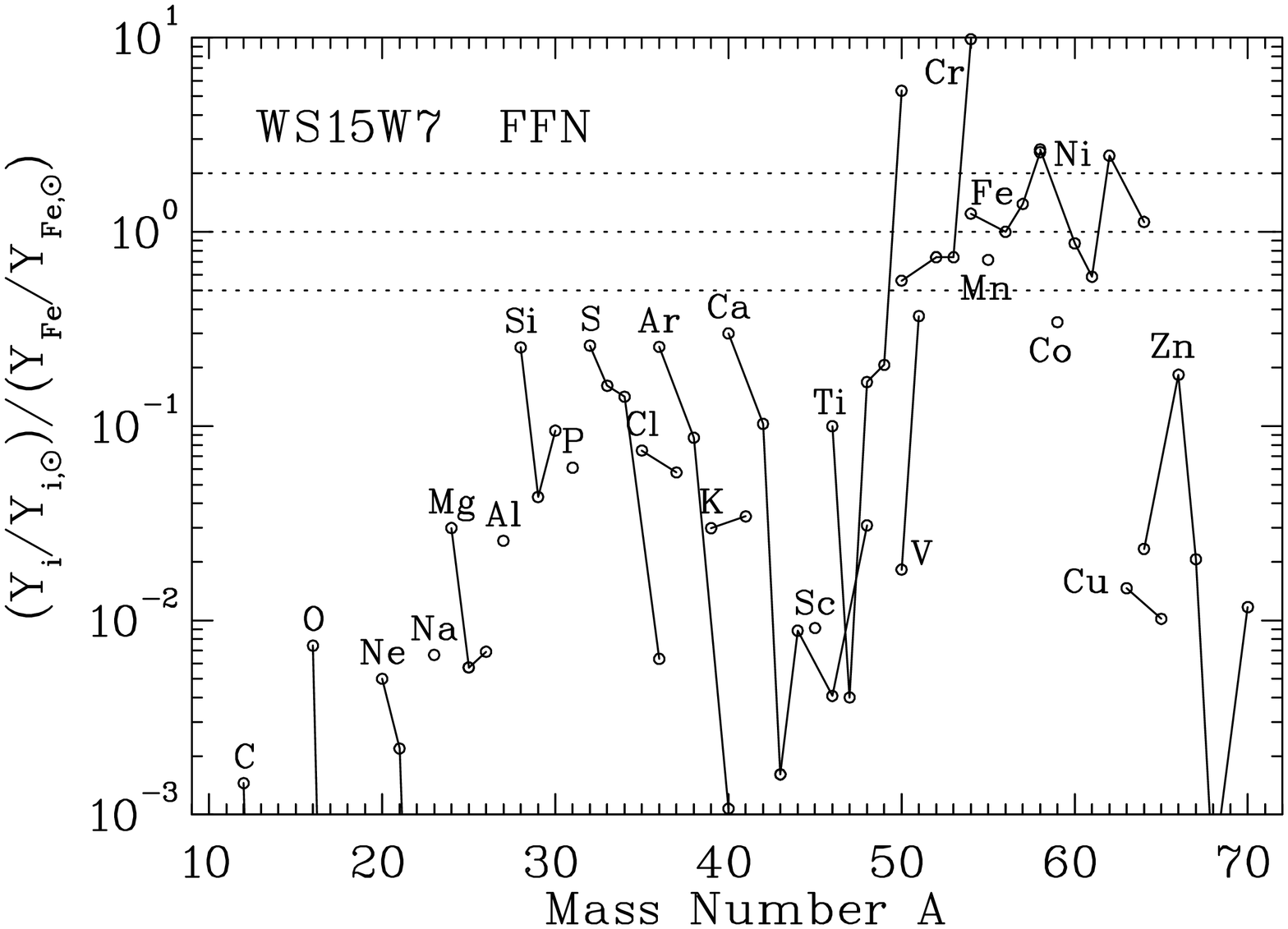}
\includegraphics[scale=0.325,angle=90]{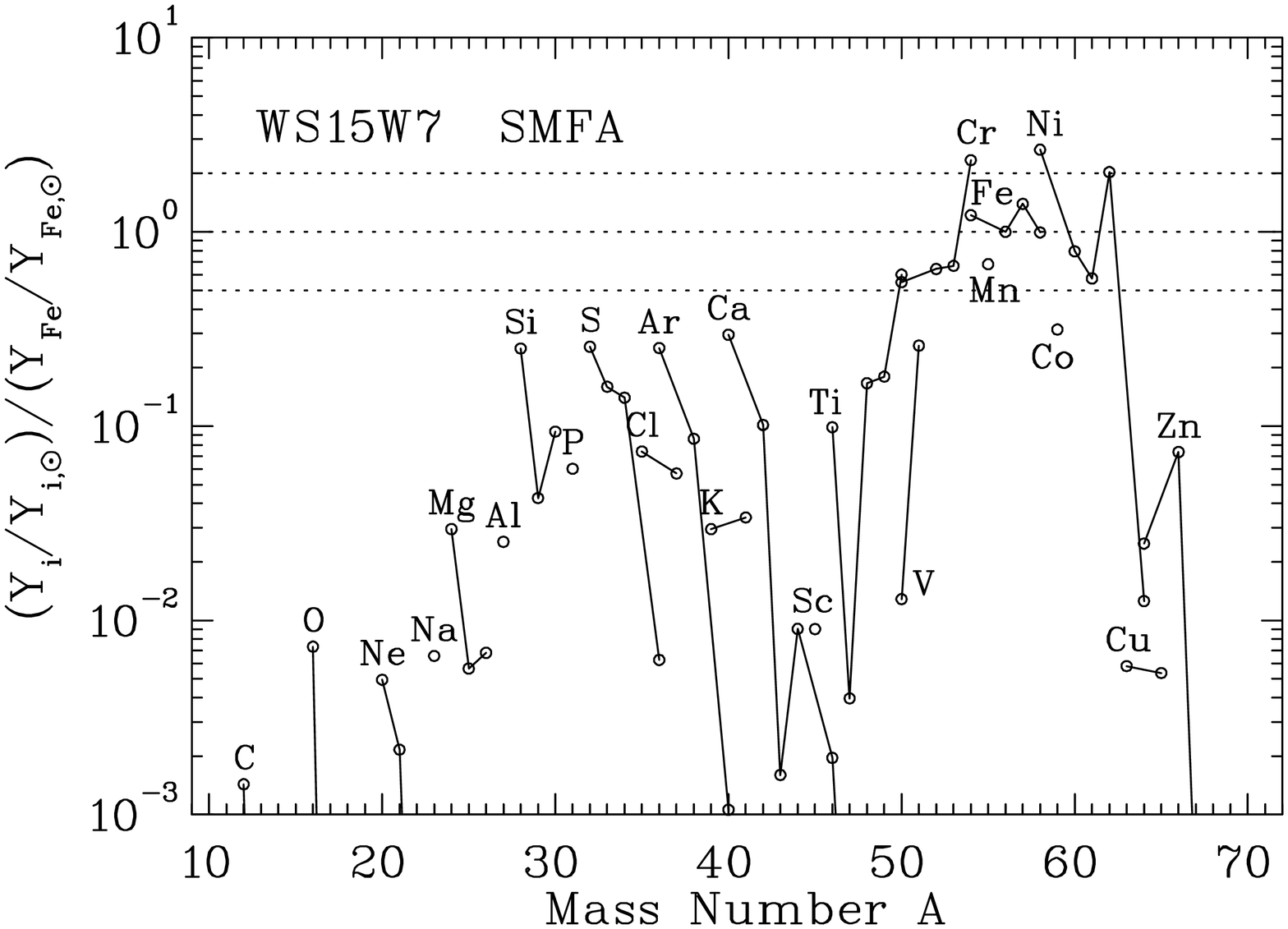}}
\end{center}
\vspace{-0.3in}
\caption{The ratio of abundances, relative to solar, as
predicted by the WS15 model. The lines connect isotopes
of one element, and the ordinate is normalized to $^{56}$Fe. 
The left panel shows results for the nucleosynthesis calculation
performed with FFN weak rates, while the right panel shows
results using scaled shell-model rates. 
}
\label{fig1}
\end{figure}

We then used these modified rates in Type Ia explosion calculations. 
The effects of the new rates on nucleosynthesis produced
by the explosions are shown in 
Fig.~\ref{fig1}. The figure shows the ratio of abundances, relative to
solar, as predicted for the WS15 model. Electron capture and 
$\beta$-decay rates were modeled with FFN rates (left panel) and with 
the modified FFN rates (SMFA, right panel). We immediately
see that the over-production of $^{54}$Cr and $^{54}$Fe
are cured by an inclusion of the interacting shell-model rates. 

\section{Conclusion}

While several astrophysical parameters are also involved
in calculations of Type Ia explosion mechanisms, 
the nuclear structure component of these reactions 
is now much more robust than it has been in the past. This
eliminates one set of parameterization and reduces the 
uncertainty of the Type Ia nucleosynthesis and explosion 
models. 
 
In the case of Type II explosions, similar and important research
is required. While some of this research is being performed presently,
it will be necessary in the near future to update all nuclear
reaction rates (electron-capture, beta-decay, neutrino scattering, 
nuclear matter opacity to neutrinos) that enter the various explosion
models and precollapse progenitors. 
Work in this direction is under way \cite{kl_pc}.
Since very neutron-rich 
nuclei become important during Type II collapse, it will become
very important to go beyond the $pf$-shell. 

While increasing computational power has moved us forward,
it is important to realize that the shell-model problem scales 
somewhat like $\exp(N)$ where $N$ is the number of valence 
particles, while 
single processor speed on high-performance machines scales as $1.8^Y$
and memory per processor scales as $1.3^Y$, where $Y$ is years. 
In 1971, Whitehead and Watt \cite{ww71}
performed the first shell-model calculations
for $^{24}$Mg in the $sd$-shell. Today we can reach nuclei
near $^{52}$Fe in a full $pf$ shell-model diagonalization using
the $M$-scheme technique. A simple analysis of the required memory
and time necessary to perform the calculation indicates that the size 
of problem in a $0\hbar\omega$ space 
that we can completely tackle with diagonalization is keeping pace
with single-processor performance and memory. One near-term
need in our field is the development of shell-model 
codes that scale in distributed-memory environments. 
When this transformation occurs, we will be able to tackle 
much more ambitious problems in very neutron-rich systems.

\end{document}